\documentclass[amsmath,amssymb,twocolumn,notitlepage,nofootinbib,superscriptaddress,floatfix,eprint,longbibliography]{revtex4-2}

\usepackage[utf8]{inputenc}
\usepackage[english]{babel}
\usepackage{graphicx}
\usepackage{color}
\usepackage[usenames,dvipsnames,table]{xcolor}
\usepackage[normalem]{ulem}
\usepackage{amsthm}
\usepackage{amsmath}
\usepackage{tikz} 
\newtheorem{lem}{Lemma}
\newtheorem{cor}[lem]{Corollary}
\usepackage[caption=false]{subfig}

\usepackage{microtype}
\microtypecontext{spacing=nonfrench}
\microtypesetup{
protrusion={true,nocompatibility},
activate={true,nocompatibility},
tracking=true,
kerning=true,
spacing={true}
}
\usepackage[all=normal,floats=tight,mathspacing=tight,wordspacing=tight,paragraphs=normal,tracking=tight,charwidths=tight,mathdisplays=normal,sections=normal,margins=normal]{savetrees}

\usepackage[style=american,autopunct=true]{csquotes}

\usepackage{placeins}
\usepackage{verbatim} 
\usepackage{mathtools}

\definecolor{blue}{rgb}{0,0,1}
\definecolor{grey}{rgb}{0.6,0.6,0.6}

\definecolor{myurlcolor}{rgb}{0,0,0.7}
\definecolor{myrefcolor}{rgb}{0.8,0,0}
\definecolor{purple}{RGB}{128,0,128}
\definecolor{ultramarine}{RGB}{63, 0, 255}
\definecolor{medblue}{RGB}{0, 0, 100}
\definecolor{googleblue}{RGB}{34, 0, 204}
\definecolor{panblue}{RGB}{0,24,150}
\definecolor{carmine}{RGB}{150, 0, 24}
\definecolor{gray}{RGB}{150, 150, 150}
\newcommand{\term}[1]{\textcolor{medblue}{\textbf{#1}}}

\newcommand{\blk}{\color{black}}

\usepackage[unicode=true,pdfusetitle, bookmarks=false,bookmarksnumbered=false,
bookmarksopen=false, breaklinks=false,pdfborder={0 0 0},backref=false,
colorlinks=true, 
linkcolor=carmine,
citecolor=googleblue,
urlcolor=panblue,
anchorcolor=OliveGreen,
]{hyperref}

\usepackage{xfrac}

\def\PR{\textsf{PR}}
\begin{document}

\title{Experimental genuine tripartite nonlocality in a quantum triangle network}

\author{Alessia Suprano}

\author{Davide Poderini}

\author{Emanuele Polino}
\author{Iris Agresti}

\affiliation{Dipartimento di Fisica - Sapienza Universit\`{a} di Roma, P.le Aldo Moro 5, I-00185 Roma, Italy}
\author{Gonzalo Carvacho}
\affiliation{Dipartimento di Fisica - Sapienza Universit\`{a} di Roma, P.le Aldo Moro 5, I-00185 Roma, Italy}
\author{Askery Canabarro}
\affiliation{International Institute of Physics, Federal University of Rio Grande do Norte, 59078-970, Natal, RN, Brazil}
\affiliation{Grupo de F\'isica da Mat\'eria Condensada, N\'ucleo de Ci\^encias Exatas - NCEx, Campus Arapiraca, Universidade Federal de Alagoas, 57309-005, Arapiraca, AL, Brazil}

\author{Elie Wolfe}
\affiliation{Perimeter Institute for Theoretical Physics, 31 Caroline St. N, Waterloo, Ontario, N2L 2Y5, Canada}

\author{Rafael Chaves}
\affiliation{International Institute of Physics, Federal University of Rio Grande do Norte, 59078-970, Natal, RN, Brazil}
\affiliation{School of Science and Technology, Federal University of Rio Grande do Norte, Natal, Brazil}
\author{Fabio Sciarrino}
\email{fabio.sciarrino@uniroma1.it}
\affiliation{Dipartimento di Fisica - Sapienza Universit\`{a} di Roma, P.le Aldo Moro 5, I-00185 Roma, Italy}

\begin{abstract}
Quantum networks are the center of many of the recent advances in quantum science, not only leading to the discovery of new properties in the foundations of quantum theory but also allowing for novel communication and cryptography protocols. It is known that networks beyond that in the paradigmatic Bell's theorem imply new and sometimes stronger forms of nonclassicality. Due to a number of practical difficulties, however, the experimental implementation of such networks remains far less explored. Going beyond what has been previously tested, here we verify the nonlocality of an experimental triangle network, consisting of three independent sources of bipartite entangled photon states interconnecting three distant parties. By performing separable measurements only and evaluating parallel chained Bell inequalities, we show that such networks can lead to a genuine form of tripartite nonlocality, where classical models are unable to mimic the quantum predictions even if some of the parties are allowed to communicate.
\end{abstract}
\maketitle
\section{Introduction}

Bell nonlocality \cite{bell1964einstein,Brunner,scarani2019bell} offers a vast research landscape with relevance for foundational \cite{Shalm,Giustina,Hensen,big2018challenging} and quantum technological applications ranging from secure quantum communication protocols \cite{gisin2007quantum,scarani2009security,pironio2013security,basset2021quantum,Chen2021} and randomness generation and certification \cite{pironio2010random,acin2016certified,bera2017randomness,agresti2020experimental,poderini2021ab} to self-testing of quantum devices \cite{vsupic2020self} and distributed computing \cite{buhrman2010nonlocality,ho2021quantum}.

Moving beyond the paradigmatic Bell scenario, the causal perspective on non-classicality \cite{wood2015lesson,chaves2015unifying,fritz2012beyond} has illuminated that new and sometimes stronger forms of nonlocality can arise \cite{renou2019genuine,gisin2017elegant,gisin2019entanglement,krivachy2019neural,baumer2021demonstrating, fritz2012beyond,henson2014theory,chaves2015information,steudel2015information,fritz2012beyond,fritz2016beyond,fraser2018causal,wolfe2019inflation,krivachy2019neural,renou2019genuine,aaberg2020semidefinite,gisin2019entanglement,kraft2020quantum,vsupic2020quantum,baumer2021demonstrating,renou2019limits,branciard2012bilocal,chaves2014causal,chaves2016polynomial,tavakoli2021bilocal,tavakoli2014nonlocal,gisin2017all,poderini2019exclusivity,tavakoli2017correlations,rosset2016nonlinear,bancal2009,chaves2017causal,gisin2019constraints,canabarro2019machine,chaves2021causal}. A particularly relevant situation is that of a network involving several parties and independent sources, a scenario akin to what one can expect from the first small to mid-sized quantum networks under development \cite{Chen2021} aiming at future large scale quantum internet \cite{wehner2018quantum,brito2020statistical}. In such networks, instead of having all distant nodes connected by a single source (as would be the case in Bell's theorem) producing a fragile many qubits state, the correlations are mediated by many sources generating states of small size and thus enhanced robustness and quality, for instance, bipartite entangled states.

In spite of significant theoretical progress, understanding the correlations that such quantum networks can give rise to (see for instance the recent review \cite{tavakoli2021bell}), the experimental side of this effort~\cite{carvacho2017experimental,saunders2017experimental,andreoli2017experimental,chaves2018quantum,polino2019device,carvacho2019perspective,sun2019experimental,poderini2020experimental,carvacho2021quantum}  is far less unexplored, due to a number of practical difficulties. In the particular case of a  photonic platform, the most prominent of these difficulties stems from the probabilistic nature of the photons' generation and the synchronization process that becomes exponentially more demanding as the number of entanglement sources increase. Furthermore, in many cases there is a need for entangled measurements, a requirement that can only be partially achieved with linear optics, or which requires device-dependent assumptions for the experimental implementation \cite{carvacho2017experimental,saunders2017experimental}. These issues are worsened when the independence of the sources is required to be physically justified, for instance, by using different and non-synchronized lasers to pump the generation crystals that act as sources \cite{poderini2020experimental}.
For this reason, only two examples of photonic networks involving independent sources have been implemented in a device-independent manner so far, one concerning the bilocality scenario \cite{carvacho2017experimental,saunders2017experimental,andreoli2017experimental,sun2019experimental,carvacho2021quantum} and the other a star network with four sources \cite{poderini2020experimental}.

\term{Genuine multipartite nonlocality} plays an important role in the context of multipartite networks.
This concept was originally introduced by~\citet{svetlichny1987distinguishing} for the intention of distinguishing those correlations whose nonclassicality does not reduce to merely bipartite nonclassicality. For Svetlichny, this distinction could only be attained by witnessing the violation of a Bell-like inequality bounding those classical models where all-but-one of the parties might communicate among themselves.
Perhaps counterintuitively, Svetlichny's notion of nonlocality does \emph{not} require multipartite quantum sources to be manifest; rather, such correlations can be realized in a network consisting of only bipartite sources~\cite{CavalcantiStar2011,AlmeidaFullyNonlocal,Contreras-Tejada2021}. Unfortunately, existing proofs of this fact rely on \emph{pure} bipartite entangled states, and hence are unsuitable for experimental demonstration.
In this work, we provide new \emph{robust} proofs of this fact by showing how the cumulative payoff of multiple bipartite nonlocal games played in parallel can effectively witness Svetlichny's genuine multipartite nonlocality.
Not only do such parallel-payoff inequalities have the advantage of being analytically derivable, they are also violable under feasible experimental conditions.
To prove this, we analyze the robustness of such parallel inequalities, thereby contrasting various multipartiteness witnesses for general network scenarios. This initial analysis lets us identify the most suitable configuration of parallel bipartite nonlocal games for experimentally verifying such multipartiteness. The optimal corresponding network structure is found to be the \term{quantum triangle}, notably a structure of extensive recent interest \cite{henson2014theory,chaves2015information,steudel2015information,fritz2012beyond,fritz2016beyond,fraser2018causal,wolfe2019inflation,krivachy2019neural,renou2019genuine,aaberg2020semidefinite,gisin2019entanglement,kraft2020quantum,vsupic2020quantum,baumer2021demonstrating,renou2019limits,gisin2017elegant}.

The triangle scenario is composed of three independent two-way sources distributing entanglement out to three parties, such that each source is shared between only two parties and such that each party receives systems from only two sources. 

Differently from the typical analysis in the literature \cite{renou2019genuine,gisin2017elegant,gisin2019entanglement,krivachy2019neural,baumer2021demonstrating, fritz2012beyond}, we here consider the situation where each of the three parties can perform a number of separable and independent measurements on each of the qubits in their possession. We experimentally realize the quantum triangle network 
using a platform based on the scheme of Ref.~\cite{poderini2020experimental}, showing that we can achieve genuine tripartite nonlocality without the (demanding!) employment of tripartite sources \cite{PironioPHD,Barrett2005}.

The paper is organized as follows. Section \ref{sec:sec2} introduces both Bell's theorem~\cite{bell1964einstein}  and Svetlichny's notion of genuine multipartite nonlocality~\cite{svetlichny1987distinguishing} from a causal perspective, highlighting how the latter can be achieved merely by playing bipartite nonlocal games in parallel. In Section \ref{sec:sec3} we present a robustness analysis of parallel nonlocal games in general bipartite networks, thereby obtaining criteria for identifying optimal scenarios and inequalities for experimental demonstration. In Section~\ref{sec:sec4} we describe our experimental implementation of the quantum triangle network, and discuss the extent to which our observed statistics witness genuine multipartiteness.

\begin{figure}[t!]
\includegraphics[width=.70\columnwidth]{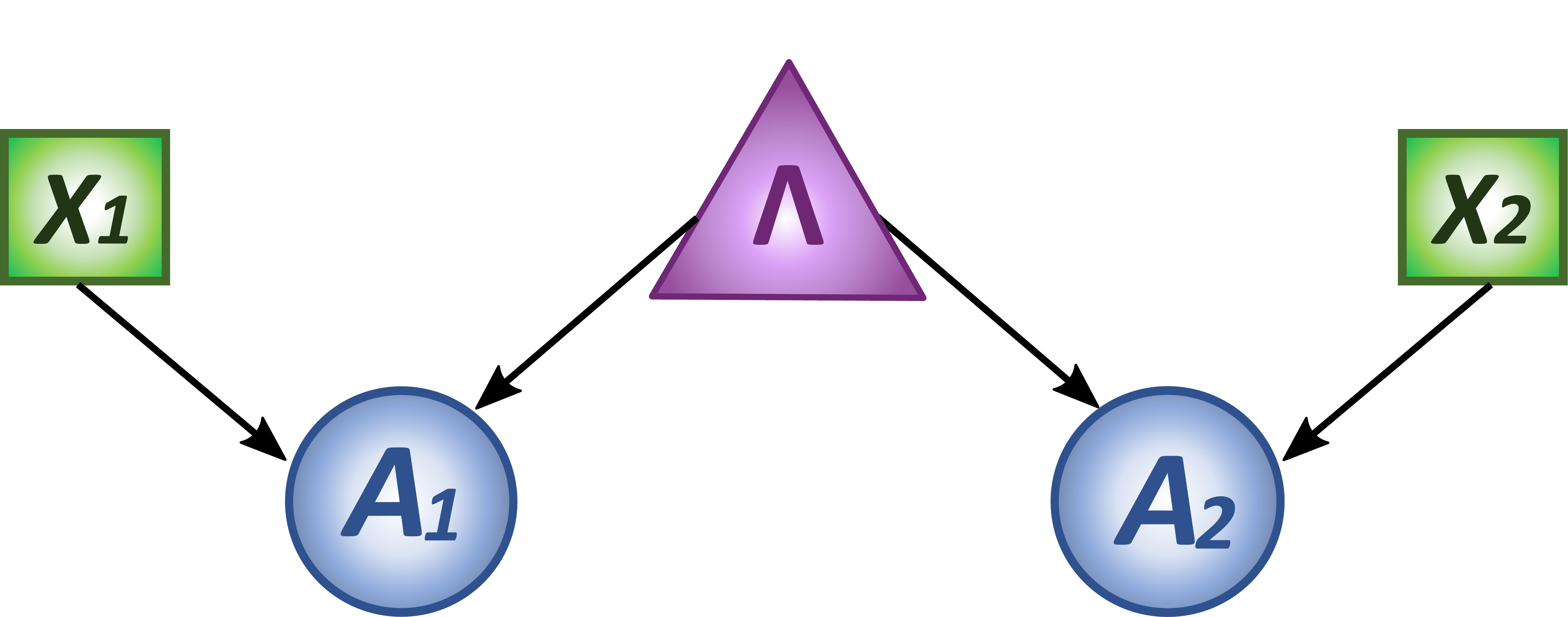}
\caption{\textbf{Directed Acyclic Graph (DAG) representation of the bipartite Bell scenario.} A classical source described by the random variable $\Lambda$ underlies the correlations observed between the measurements outcomes of two distant parties.}
\label{fig:dag1} 
\end{figure}

\section{Bell's theorem and causal structures}
\label{sec:sec2}
Bell's theorem~\cite{bell1964einstein} is typically cast as the incompatibility of quantum predictions with those expected from theories respecting local causality, the latter formally expressed by local hidden variables (LHV) models. More specifically, two distant parties, Alice and Bob, receive their shares of a physical system produced by a source that classically is represented by a random variable $\lambda$. Since we cannot assume to have access to all relevant degrees of freedom of the source, $\lambda$ is treated as a hidden variable, information about which we can only gather via measurements. Upon receiving their part of the joint physical system, the parties randomly decide which kind of measurement to perform, a choice parametrized by the variables $x_1$ and $x_2$ with corresponding outcomes $a_1$ and $a_2$ for Alice and Bob, respectively. Any observable probability distribution $p(a_1,a_2\vert x_1,x_2)$ compatible with the LHV assumptions can then be decomposed as
\begin{equation}
\label{eq:LHV}
    p(a_1,a_2\vert x_1,x_2)= \sum_{\lambda}p(a_1\vert x_1,\lambda)p(a_2\vert x_2,\lambda)p(\lambda),
\end{equation}
where we have explicitly employed the causal assumptions in Bell's theorem. First, the locality assumption states that the outcome of Alice (similarly for Bob) should only depend on the variables on her causal past (thus the necessity of space-like separation between Alice and Bob). It implies that $p(a_1\vert x_1,x_2,a_2,\lambda)=p(a_1\vert x_1,\lambda)$ (similarly for Bob). In turn, the measurement independence assumption (also known as free-will) states that $p(x_1,x_2,\lambda)=p(x_1,x_2)p(\lambda)$ and encompasses the basic idea that the observers should have the freedom to choose which observable to measure independently of how the system under measurement is prepared. Implicit is the realism assumption, represented by the fact that the LHV model assumes the existence of the variable $\lambda$ subsuming well-defined probabilities for all possible outcomes, even for those measurements that are not performed.

From a more modern perspective, we can see Bell's theorem as the fact that a classical causal model cannot reproduce the quantum correlations, that according to Born's rule are given by 
\begin{equation}
    p(a_1,a_2\vert x_1,x_2)= \mathrm{Tr}\left[(M_{a_1}^{x_1} \otimes M_{a_2}^{x_2}) \rho \right],
\end{equation}
where $M_{a_1}^{x_1}$ are POVM operators describing Alice's measurements (similarly for Bob) and $\rho$ is the density operator representing the physical system shared between Alice and Bob. We are imposing a given causal structure to the experiment, that classically is represented via a direct acyclic graph (DAG) in Fig. \ref{fig:dag1}. From the causal Markov condition \cite{pearl2009causality} -- implying that a given node of the graph should be independent of all its non-descendants given its parents-- we obtain exactly the LHV decomposition \eqref{eq:LHV}. That is, in order to explain quantum correlations we have to consider a quantum version of a causal model \cite{henson2014theory,chaves2015information,fritz2016beyond,allen2017quantum}.

\begin{figure}[t!]
\begin{minipage}[b]{0.45\columnwidth}
\subfloat[Svetlichny scenario]{\includegraphics[width=0.85\textwidth]{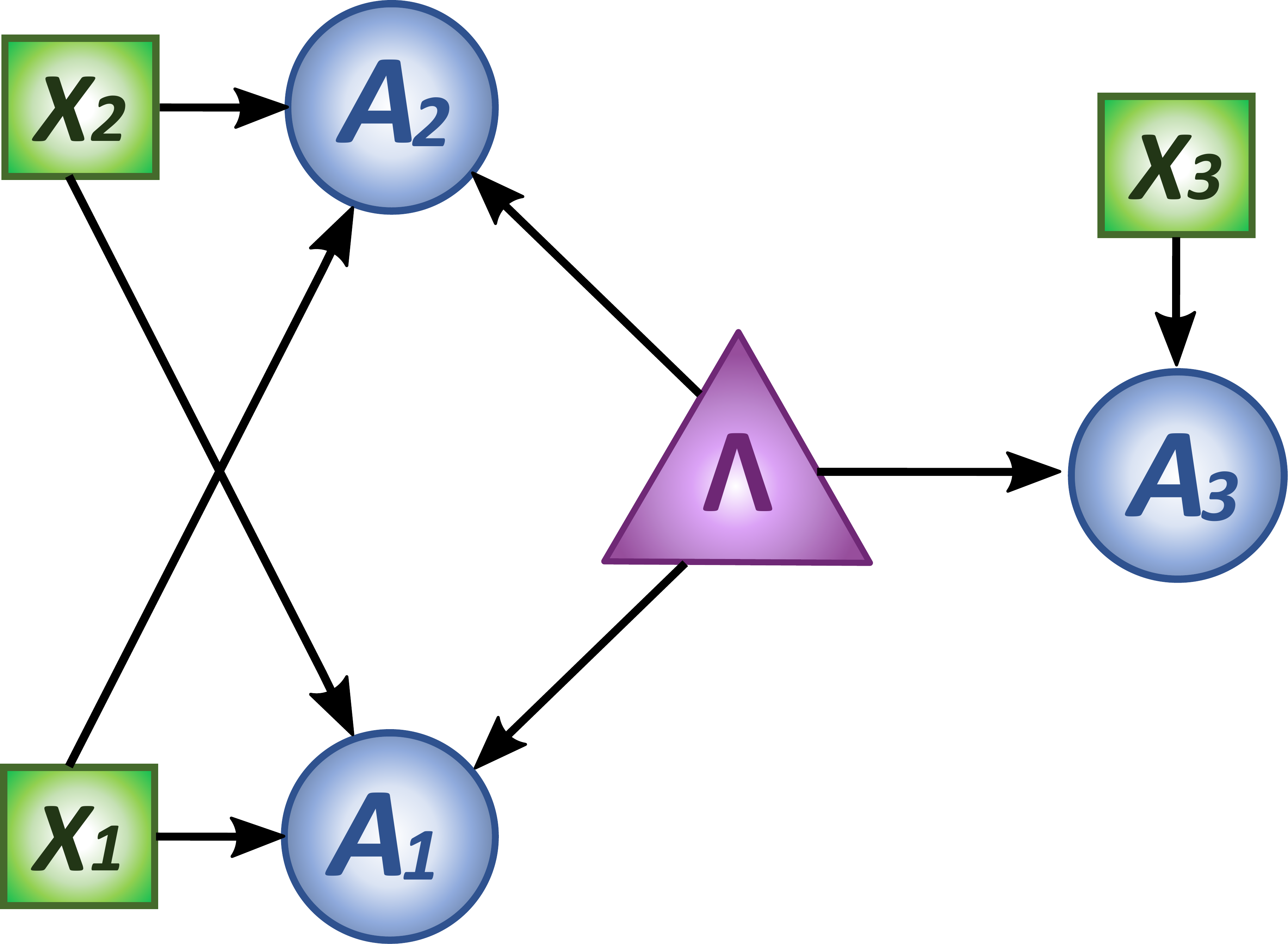}
    \label{fig:Svetlichny_scenario}}
\end{minipage}
\begin{minipage}[b]{0.45\columnwidth}
\centering
\subfloat[Quantum triangle scenario]{\includegraphics[width=0.95\textwidth]{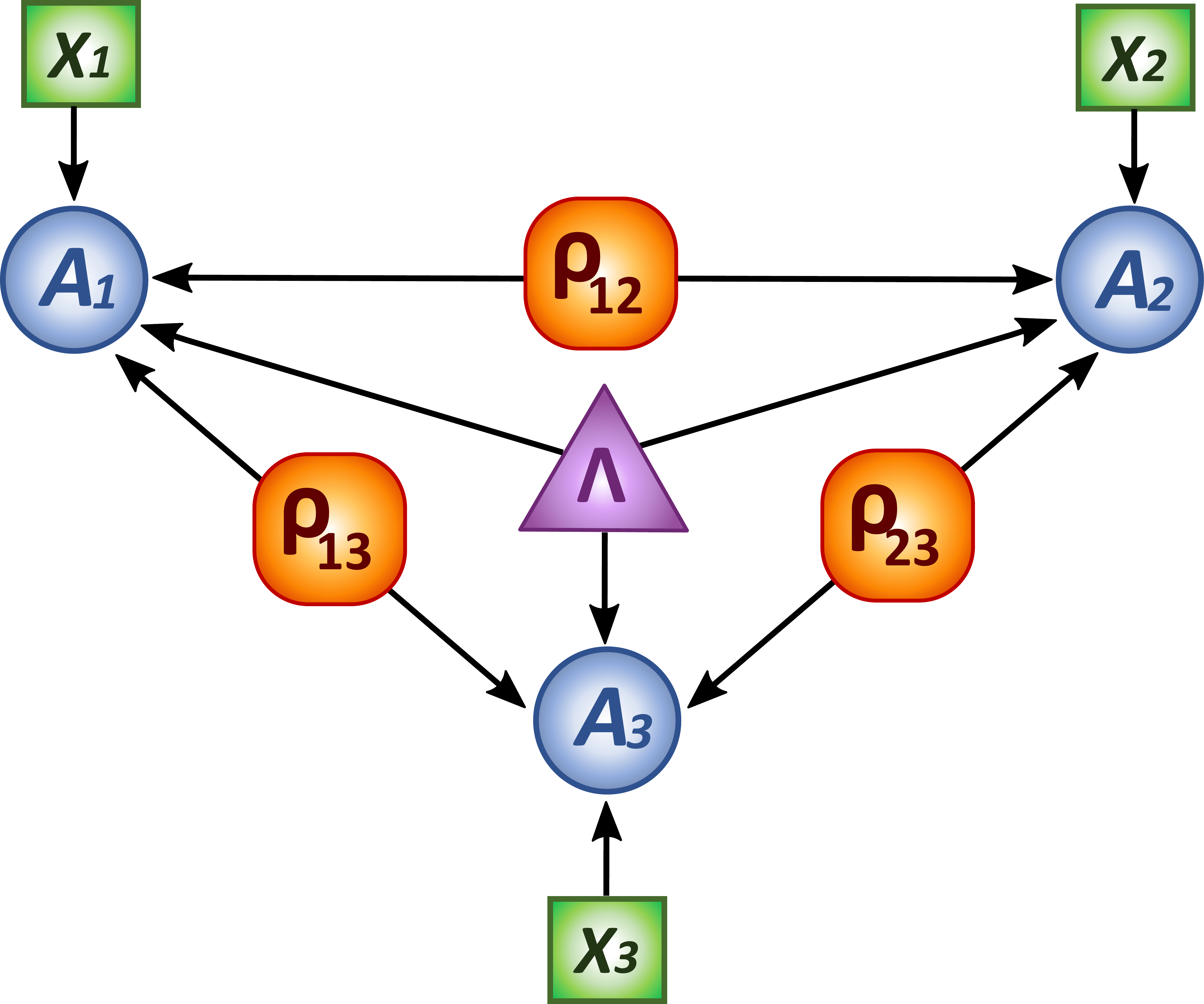}
    \label{fig:triangle_scenario}}
\end{minipage}
\caption[.]{\textbf{Directed Acyclic Graph (DAG) representations of tripartite networks.} \subref{fig:Svetlichny_scenario} In the tripartite Svetlichny scenario --- i.e., in the causal structure of Svetlichny's classical foil theory --- a classical source described by the random variable $\Lambda$ underlies the correlations observed between the measurements outcomes of the three parties. In contrast to Bell's local hidden variable models per Fig~\ref{fig:dag1}, however, in Svetlichny's scenario any two of the parties can communicate, WLOG, by exchanging their inputs. \emph{Which} pair of parties are in communication may also depend on $\Lambda$, though this additional freedom is not explicitly depicted here.  
\subref{fig:triangle_scenario} A conceptual scheme of the apparatus realizing a quantum triangle network. Note the three independent quantum sources $\rho_{12}$, $\rho_{13}$ and $\rho_{23}$ that are shared between the different pairings of the three parties. In practice, the classical common cause $\Lambda$ is only used to establish a common reference frame.
Our main result is that by allowing the non-central sources in (b) to be quantum, we can experimentally achieve correlations which cannot be explained in (a).}
\end{figure}

A straightforward generalization of Bell's theorem is to consider the multipartite scenario where all $n$ distant parties are connected by a single source \cite{Brunner}. Classically, the observed correlation should thus be decomposable as
\begin{align*}
    p(a_1,...,a_n\vert x_1,...,x_n)= \textstyle\sum\limits_{\lambda}p(a_1\vert x_1,\lambda)... p(a_n\vert x_n,\lambda)p(\lambda).
\end{align*}
Just as in the bipartite case, allowing for the source to be quantum can give rise to distributions strictly outside the set of multipartite LHV distributions. This may seem trivial at first: if quantum violation can be achieved in the bipartite scenario, surely quantum violation can be achieved in the multipartite scenario! One needs only to consider \emph{bipartite marginals} of a tripartite scenario to observe quantum advantage. What makes the multipartite scenario interesting, then, is that the sorts of operational advantages afforded by a tripartite quantum source are qualitatively distinct from the more limited advantages afforded by bipartite quantum sources. Early pioneers of the study of multipartite nonlocality were therefore motivated to define a notion of \emph{genuine multipartiteness} of a nonclassical correlation. The original definition for genuine multipartiteness was proffered by Svetlichny in 1987~\cite{svetlichny1987distinguishing}. As with all subsequent definitions of genuine multipartiteness, Svetlichny's definition was apophatic. That is, a distribution obtained by quantum strategies is said to be genuinely mutlipartite if it \emph{cannot} be explained by a suitable foil theory. The foil theory that Svetlichny envision is entirely classical, but wherein all but one of the involved parties can communicate with each other~\cite{svetlichny1987distinguishing,bancal2009,chaves2017causal}, and wherein the strict subset of parties which may communicate need not be fixed, but can depend on a hidden variable.

For instance, in the tripartite scenario we can consider a convex combination of models where any two of the parties can communicate arbitrarily among them, a situation that can be without loss of generality \cite{chaves2017causal} be described by the communication of inputs between the parties, with the corresponding DAG shown in Fig.~\ref{fig:Svetlichny_scenario} for the case where parties 1 and 2 are the communicating ones. The classical description of the tripartite probability distribution is thus given by
\begin{align}
    p(a_1,a_2,a_3 &\vert x_1,x_2,x_3)= q_{1}p_{a_1 \leftrightarrow a_2}+q_{2}p_{a_1 \leftrightarrow a_3}+q_{3}p_{a_2 \leftrightarrow a_3},\nonumber\\
&\text{where } q_1+q_2+q_3=1,\text{ and} \nonumber\\\label{eq:svet_model1}
    p_{a_1 \leftrightarrow a_2}=& \sum_{\lambda}p(a_1 \vert x_1,x_2,\lambda)p(a_2 \vert x_1,x_2,\lambda)p(a_3 \vert x_3,\lambda)
\end{align}
is the nonlocal classical model represented by the DAG in Fig. \ref{fig:Svetlichny_scenario} (similarly for the terms $ p_{a_1 \leftrightarrow a_3}$ and $ p_{a_2 \leftrightarrow a_3}$). Hereafter we abbreviate the oft-referred to concept of Svetlichny-esque genuine multipartite nonlocality as \term{SGMNL}.

Achieving SGMNL correlations would require entanglement \emph{beyond} the sort that can be achieved from scratch with local operations and classical communication (LOCC) and a quantum channel connecting only two of the three parties in the network. For this reason, Svetlichny's definition for genuine multipartite nonlocality serves as a device-independent witness of genuine multipartite entanglement.

Intuitively, one might think that Svetlichny's definition is meant to distinguish certain correlations realized by a 3-way-quantum source from those that can be realized from 2-way-quantum sources, but it is incorrect. SGMNL is tailored to witness \emph{nonrealizability by LOCC and quantum channels only between a strict subset of the parties}. As such, Sveltichny's definition can be \emph{hacked}: Using only local operations on bipartite quantum sources --- without even making use of shared randomness or even classical communication between strict subsets of parties --- one can realize SGMNL correlations. The causal structure  of this so-called ``triangle" scenario is depicted in Fig.~\ref{fig:triangle_scenario}, such that a \emph{quantum} distribution which can be realized in  such scenario is of the form 
\begin{align}
\begin{split}
p^{\mathrm{Q}}(&a_1,a_2,a_3\vert x_1,x_2,x_3)=\\
&\mathrm{Tr}\left(\rho_{12} \otimes \rho_{13} \otimes \rho_{23}\,\cdot\, M_{x_1}^{a_1}\otimes M_{x_2}^{a_2} \otimes M_{x_3}^{a_3}\right),    
\end{split}
\end{align}
where $\rho_{ij}$ represents the bipartite quantum state generated by the source $\Lambda_{ij}$ and $M_{x_i}^{a_i}$ are the opportunely ordered  positive-operator valued measures (POVMs) associate to node $i$, such that $\sum_{a_i}M_{x_i}^{a_i}=\openone$, for all $x_i$.

The main result of this article is an experimental demonstration of the achievement of SGMNL using a quantum triangle causal structure. In fact, we further tie our hands behind our backs and showcase Svetlichny-esque multipartiteness from \emph{product} measurements on bipartite sources. 
At this point, it is worthy remarking an issue present in most experimental implementations of quantum networks but that nevertheless has no impact in our experiment. 
While in usual Bell test resorting to a single source of correlations, shared reference frames (used to establish the common frame for qubit measurements, for instance) do not change the causal structure under test, this is no longer true for networks involving independent sources \cite{andreoli2017experimental}. For instance, if in the triangle one employs a common reference frame, this could be used to establish hidden correlations between parties that are assumed to be independent and thus hinder the possibility of witnessing non-classicality. As we are testing quantum correlations against the classical Svetlichny model in Fig. \ref{fig:Svetlichny_scenario}, shared reference frames are permitted, reason why we explicitly included it in the DAG description of our triangle experiment in Fig. \ref{fig:triangle_scenario}.

Although multiple prior works have recognized the susceptibility in Svetlichny's definition discussed above, all prior proofs are remarkably unsuitable for experimental implementation. Accordingly, we briefly review some relevant theory results, but then we derive novel inequalities for witnessing SGMNL which are amenable for experimental analysis.

In Refs.~\cite{PironioPHD,BarrettPironio2005} it is shown that local wirings of extremal bipartite nonsignalling boxes leads to SGMNL. Local wirings are a form of local operations, and this does constitute a triangle-scenario causal structure achieving SGMNL, but the protocols in Refs.~\cite{PironioPHD,BarrettPironio2005} rely on post-quantum resources, and hence their ideas cannot be adapted for quantum experimentation.
Refs.~\cite{CavalcantiStar2011} proves that entanglement swapping can asymptotically lead to steering to a maximally entangled pair of qubits on any hub-and-spoke node pair in the star network, which implies SGMNL per the results of Ref.~\cite{AlmeidaFullyNonlocal}. Our goal is to use product measurements, however, and anyway asymptotics are not amenable to experimentation. Nevertheless Refs.~\cite{CavalcantiStar2011,AlmeidaFullyNonlocal} prove that SGMNL can arise in-principle from local operations of parties in a network defined by merely-bipartite quantum states. Ref.~\cite{Contreras-Tejada2021} takes this one step further, and shows that any star network composed of bipartite entangled pure states admits local measurements which give rise to SGMNL correlations. From an experimental perspective, Ref.~\cite{Contreras-Tejada2021}'s proof that SGMNL can be achieved without \emph{maximal} entanglement is appreciated, but since that proof continues to rely on \emph{pure} states and \emph{noiseless} measurements, it too is not amenable for experimental investigation.

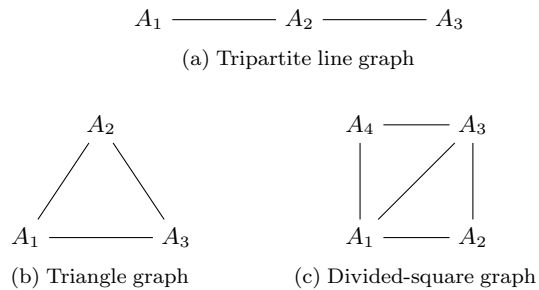
\begin{figure}[t]
\subfloat[Tripartite line graph]{    \begin{tikzpicture}
         \node (a1) at (0,0) {$A_1$};
         \node (a2) at (2,0) {$A_2$};
         \node (a3) at (4,0) {$A_3$};
         \path (a1) edge (a2) (a2) edge (a3);
    \end{tikzpicture}
    
    \label{fig:line_graph}}
\\
\bigskip
\begin{minipage}[b]{.35\columnwidth}
\subfloat[Triangle graph]{    \begin{tikzpicture}
         \node (a1) at (0,0) {$A_1$};
         \node (a2) at (1,1.5) {$A_2$};
         \node (a3) at (2,0) {$A_3$};
         \path (a1) edge (a2) (a2) edge (a3) (a1) edge (a3);
    \end{tikzpicture}
   
    \label{fig:triangle_graph}}
\end{minipage}
\begin{minipage}[b]{.6\columnwidth}

\subfloat[Divided-square graph]{    \makebox[5cm][c]{\begin{tikzpicture}
         \node (a1) at (0,0) {$A_1$};
         \node (a2) at (1.5,0) {$A_2$};
         \node (a3) at (1.5, 1.5) {$A_3$};
         \node (a4) at (0,1.5) {$A_4$};
         \path (a1) edge (a2) (a1) edge (a4) (a2) edge (a3) (a3) edge (a4) (a3) edge (a1);
    \end{tikzpicture}}
    
\label{fig:square_graph}}
\end{minipage}
\caption[]{
\textbf{Graph representation of nonlocal games.} We can represent a particular configuration of a nonlocal game, composed by parallel bipartite ones, using an undirected graph, where each node represents a party and two nodes are connected by an edge if the corresponding bipartite game appears in the payoff function. 
Above we show two instances of possible configurations for tripartite games; the tripartite line \subref{fig:line_graph} consisting of two parallel bipartite games, and the triangle \subref{fig:triangle_graph} consisting of three parallel bipartite games. We also depict a quadpartite scenario \subref{fig:square_graph} consisting of five parallel bipartite games.}\label{fig:graph}
\end{figure}

Our perspective is to consider \emph{parallel nonlocal games} such as it might be implemented in the triangle-like scenario of Fig.~\ref{fig:triangle_graph} or in more general scenarios.  In general, we can model such a network as a graph, with nodes corresponding to parties and edges corresponding to a pair of parties playing an instance of a nonlocal game. We can then ask about the \emph{maximum total payoff} of all parallel nonlocal games if the correlations are limited to those explainable by Svetlichny's foil theory. For instance, the classical tripartite correlations that can be explained by the DAG in Fig.~\ref{fig:Svetlichny_scenario} are such that, for three nonlocal games being played in parallel in the triangle scenario, the total score would have to satisfy
\begin{equation}
\label{eq:svet_bi}
I^{12}+I^{13}+I^{23}\leq 2B_{L}+B_{S},  
\end{equation}
where $I_{ij}$ is a generic bipartite Bell inequality between parties $i$ and $j$ given by
\begin{equation}
   I_{ij}=\sum \alpha_{a_{ij},a_{ji},x_i,x_j}p(a_{ij},a_{ji}\vert x_i,x_j) \leq B_L, 
\end{equation}
where $a_{ij}$ is the measurement outcome related to the measurement performed by party $i$ on its state shared with party $j$. A given outcome $a_i$ in the triangle can be understood as the composition of such individual outcomes, for instance, $a_1=(a_{12},a_{13})$. Continuing, $B_L$ is the local bound achievable by LHV models of the form \eqref{eq:LHV} and $B_{S}$ is the algebraic maximum that can be obtained by a NLHV model where both parties can communicate with each other. The proof of this inequality follows from the analysis in \cite{chaves2017causal} and can be intuitively understood as follows. If the parties $1$ and $2$ can communicate then they can achieve the bound $B_S$, however, since the parties $1$ and $3$ as well as $2$ and $3$ cannot communicate, they are indeed limited by the local bound $B_L$. In particular, notice that the same argument holds for any permutation of which two parties can communicate, thus leading to the Bell inequality \eqref{eq:svet_bi} bounding Svetlichny's model \eqref{eq:svet_model1}. 

\section{Optimizing a Quantum Experiment to achieve SGMNL}\label{sec:sec3}
Let a graph define a sum of bipartite Bell inequalities in the manner described above. That is, the triangle graph (Fig.~\ref{fig:triangle_graph}) corresponds to ${I^{12}+I^{13}+I^{23}}$, the tripartite line graph (\ref{fig:line_graph}) to ${I^{12}+I^{23}}$, and the divided-square graph (\ref{fig:square_graph}) 
to ${I^{12}+I^{13}+I^{23}+I^{14}+I^{34}}$. 

\begin{lem}[Total Payoff within Svetlichny's Foil Theory]\label{ref:dibound}
The total score associated with a given graph $G$ over correlations that are \emph{not} SGMNL is upper bounded by
\begin{align*}
&\min_{v\in\text{vertices}_G} \left(\begin{array}{l}
\hphantom{+}B_{L}\times [\text{\# of edges connected to v}] 
\\ + B_{S}\times [\text{\# of edges not connected to v}] \end{array}\right)\\
=\quad& B_{S}\times [\text{\# of edges of }G] 
\\- &(B_{S}-B_L)\min_{v\in\text{vertices}_G} [\text{\# of edges connected to v}] 
\end{align*}
\end{lem}
On the other hand, the total score under quantum strategies is just the quantum bound times the number of edges.
\begin{lem}[Total Payoff with Bipartite Quantum Strategies]
The total score associated with a given graph $G$ over correlations achieved by playing quantumly-realizable bipartite nonlocal games in parallel is
\begin{align*}
B_{Q}\times [\text{\# of edges of }G],
\shortintertext{or, for a imperfect realizations,}
\left(v\, B_{Q}+(1-v)B_N\right)\times [\text{\# of edges of }G]
\end{align*}
\end{lem}
where $B_N$ is the payoff score of the nonlocal game under white noise. For all the Bell inequalities we will be considering we have $B_N=0$, which makes the noise threshold for quantum achievement of SGMNL quite straightforward.
\begin{cor}[Visibility for SGMNL via Bipartite Quantum Strategies]\label{cor:critvis}
If an experiment can achieve quantum correlations with visibility $v$ and $B_N=0$, then playing such correlation in parallel according to graph $G$ yields SGMNL whenever
\begin{align*}
v > \frac{B_S}{B_Q} - \frac{(B_{S}-B_L)\displaystyle\min\limits_{v\in\text{vertices}_G} [\text{\# of edges connected to v}]}{B_{Q}\times [\text{\# of edges of }G]} 
\end{align*}
\end{cor}

Thus, to achieve SGMNL using parallel bipartite strategies one should optimize both the \emph{particular Bell inequality} (bipartite nonlocal game) corresponding the every edge in the graph, and also one should optimize the \emph{graph itself} to ensure that the ratio of $B_S$-scoring games to $B_L$ scoring games is as small as possible.

\subsection{Optimizing the Bell Inequality}
Before invoking multisetting generalizations, let us consider the CHSH inequality. This inequality notably has $B_S=4$, $B_L=2$, $B_Q=2\sqrt{2}$, and $B_N=0$. Does this lead to SGMNL? In some graphs yes, in other graphs no.
For instance, in the tripartite-line graph, the required $v$ per Cor.~\ref{cor:critvis} for CHSH would exceed unity, in other words, parallel CHSH with quantumly-realizable scores \emph{cannot} be used to witness SGMNL in the tripartite line scenario.

At first, one might think that this could  be an artifact of the Cor.~\ref{cor:critvis} being a \emph{sufficient but not necessary} criterion for witnessing SGMNL. However, we can \emph{explicitly} reconstruct the correlations associated with parallel quantum CHSH in the tripartite-line graph via a convex mixture of the extremal correlations realizable in Fig.~\ref{fig:Svetlichny_scenario} (and relabeling thereof). To see this, let $P^{ij}_{PR}$ denote a PR box between parties $i$ and $j$, which yields a CHSH score of $+4$. 
Let $P^{ij}_{\overline{\PR}}$ denote the anti-PR box between parties $i$ and $j$, which yields a CHSH score of $-4$. 
Let $\PR^{ij}(v)\coloneqq v\times P^{ij}_{\PR} + (1-v)\times P^{ij}_{\overline{\PR}}$. 
In this notation, the Tsirelson box~\cite{cirel1980quantum} which gives CHSH=$2\sqrt{2}$ is given by $P_{Tsirelson}^{ij}=\PR^{ij}(\frac{2+\sqrt{2}}{4})$. 
The reader may verify that \begin{align}\nonumber
&\hspace{-3ex}P_{Tsirelson}^{12}\otimes P_{Tsirelson}^{23}= 
\\ \Bigg( &{w \frac{\PR^{12}(1)\otimes \PR^{23}(\sfrac{3}{4})+\PR^{12}(\sfrac{3}{4})\otimes \PR^{23}(1)}{2}} 
\\ &+ {{(1-w)}\frac{\PR^{12}(0)\otimes \PR^{23}(\sfrac{1}{4})+\PR^{12}(\sfrac{1}{4})\otimes \PR^{23}(0)}{2}} \Bigg),\nonumber
\end{align}
where $w=\frac{3+2\sqrt{2}}{6}\approx 0.971$. This provides an explicit convex decomposition of parallel Tsirelson boxes in terms of four boxes which are themselves clearly members of the Svetlichny polytopes (i.e. the sets of correlations achievable with classical communication between two of the three parties).

On the other hand, the CHSH inequality \emph{does} lead to SGMNL for the \emph{triangle} graph (Fig.~\ref{fig:triangle_graph}), as Cor.~\ref{cor:critvis} tells us that playing CHSH in parallel per the triangle graph witness SGMNL down to a visibility of $v\approx 0.943$.

Note that this clarifies the potentially misleading statement in Ref.~\cite{coiteux2021no} which stated that ``the correlations obtained from CHSH violations in parallel between Alice and Bob as well as between Bob and Charlie fulfill Svetlichny’s criterion for genuine tripartite nonlocality." That is true algebraically, as parallel \emph{PR} boxes in the tripartite line scenario (Fig.~\ref{fig:line_graph}) are SGMNL, but parallel \emph{Tsirelson} boxes are \emph{not} SGMNL in the tripartite line scenario, though they are SGMNL in the triangle scenario (\ref{fig:triangle_graph}).

We are able to improve the minimum visibility, however, by considering inequalities where $B_Q$ is even closer to $B_S$ than to $B_L$. In particular, we consider the ``chained" or ``Barrett-Kent-Pironio (BKP)" family of Bell inequalities~\cite{braunstein1990wringing,BKP} which generalize CHSH to the case of more setting. 
For our purposes we express the BKP game between parties $i$ and $j$ as
\begin{equation}
        S_k^{ij} = \sum_{l=1}^{k} 
        \left[
        \left< A^i_l A^j_l \right> +\left< A^i_{l+1} A^j_{l} \right>
        \right]
        \leq 2k-2 \\
    \label{eq:chained}
\end{equation}
with $A^{i}_{k+1} = -A^{i}_1$ and $\left< A^i_l A^j_{l^{\prime}} \right>$ is the expectation value when the corresponding observables $A^i_l=\sum_{a_i} a_i M^{a_i}_{x_i=l}$ are measured by the distant parties. The BKP family of nonlocal games have the properties of
\begin{align}
B_L = 2 k -2, \;\;\;\; B_S = 2 k, \;\;\;\; B_Q = 2 k \cos\left(\frac{\pi}{2 k}\right), \;\;\;\; B_N=0,
\end{align}
where $k$ denotes the number of settings available to each party.
The BKP games are noteworthy for having ${B_Q \to B_S}$ as ${k\to \infty}$~\cite{BKP}, and for recovering the CHSH game when $k{=}2$. Does that mean that more settings always improves the critical visibility for achieve SGMNL with parallel BKP games? No! In the limit ${k\to \infty}$ we also have $B_L/B_Q$ approach 1. Thus, every graph requires its own optimization.
For the tripartite-line graph we can witness SGMNL with $v>0.947$ at $k{=}5$. For the triangle, however, the optimum number of settings is $k{=}3$, which witnesses SGMNL whenever $v>0.899$. 

Is it always the case that playing bipartite games in parallel per the triangle graph (Fig.~\ref{fig:triangle_graph}) collectively requires less visibility to witness SGMNL than playing bipartite games in parallel per the tripartite line graph (\ref{fig:line_graph})? Yes, as we next discuss.

\subsection{Optimizing the Graph}
Why does the triangle graph outperforms the line graph? The answer is that in the triangle graph no node as a vertex degree less than average. The total number of edges is equal to the number of nodes times the average vertex degree divided by two. By inspection of Cor.~\ref{cor:critvis} it is clear that optimal graphs will have each node connected to the same number of edges. The technical term for such graphs is that they are \term{regular}. The triangle graph is regular; the tripartite line graph is not.
Let's take a moment to compute a visibility such that a regular graph will witness SGMNL per Cor.~\ref{cor:critvis}. If $[\text{\# of edges connected to }v]=d$ for all $v$, such that $[\text{\# of edges of }G]=\frac{d}{2}\times[\text{\# of nodes of }G]$ then we have
\begin{align}
v > \frac{1}{B_Q}\left(B_S-\frac{2\times(B_S-B_L)}{[\text{\# of nodes of }G]}\right)\,.
\end{align}
This has two important consequences. Firstly: the optimal network to witness SGMNL for parallel quantum nonlocal games is the triangle. Going to four or more parties only increases the requisate visibility. Secondly, achieving Svetlichny-esque \emph{fully multipartite} nonlocal correlation \emph{for any number of parties} $N$ can evidently be realized by parallel playing suitably-optimized BPK games in a regular graph with $N$ nodes. Indeed, if we fix $k{=}N$ we find that
$v > \frac{N^2-2}{N^2 \cos(\frac{\pi}{2N})}$\,,
which is strictly less than one for all integer-valued $N\geq 3$.

We proceeded, therefore, to implement parallel BKP games in the quantum triangle scenario with multiple settings per party. 
\blk 

\section{Experimental genuine tripartite nonlocality in the triangle network}
\label{sec:sec4}

\begin{figure}[t!]
\includegraphics[width=\columnwidth]{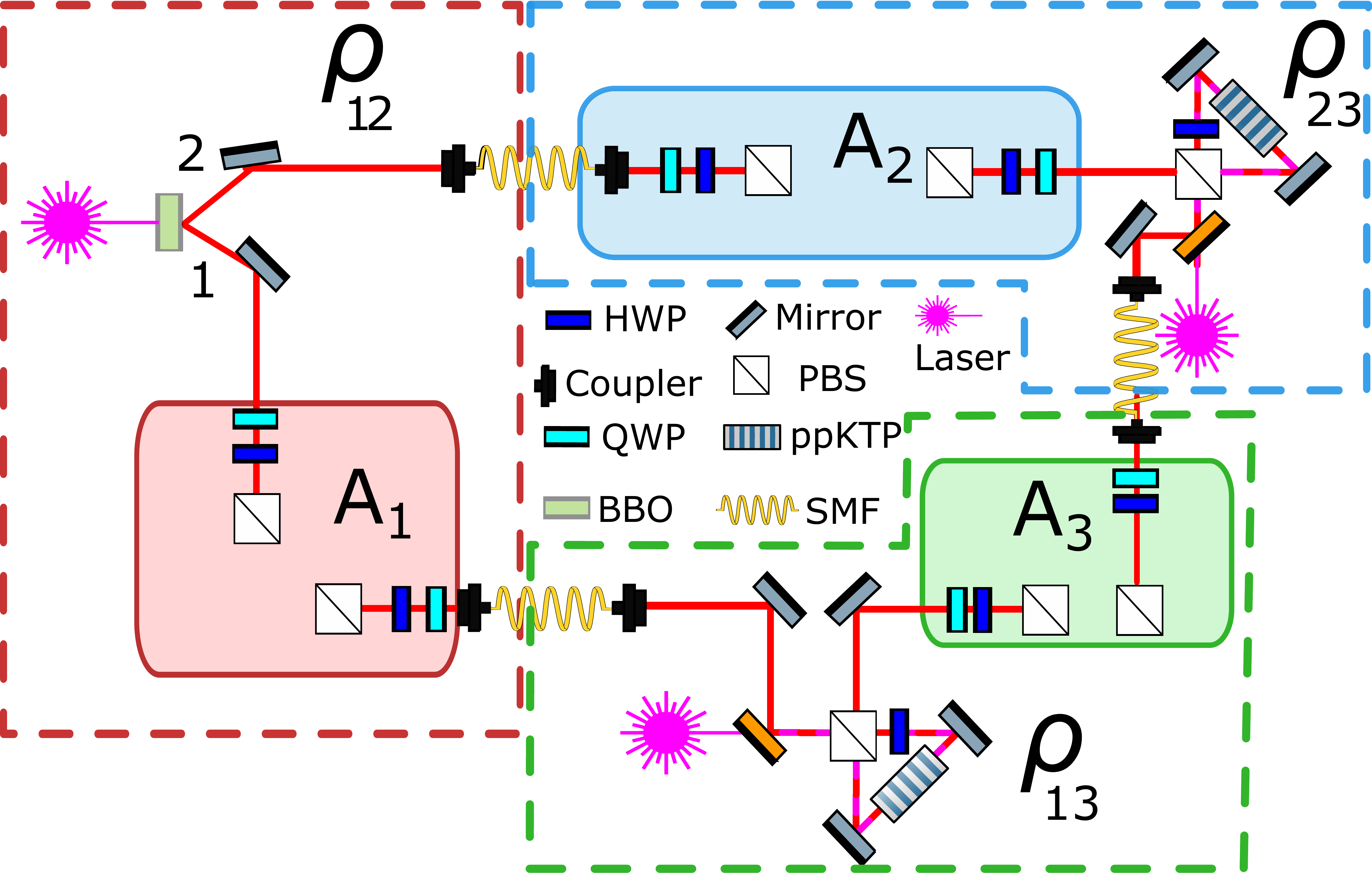}
\caption{\textbf{ Experimental apparatus.} Three independent  sources, $\rho_{12}$, $\rho_{13}$ and $\rho_{23}$, generate polarization-entangled photon pairs. A pump laser for source $\rho_{12}$  with wavelength $\lambda = 397.5$ nm is produced by a second harmonic generation (SHG) process from a Ti:Sapphire mode locked laser with repetition rate of $76$ MHz, and is focused on a 2 mm-thick beta-barium borate (BBO) crystal. Sources $\rho_{13}$ and $\rho_{23}$ employ a continuous-wave diode laser, one for each source, with wavelength $\lambda = 404$ nm which pumps a 20mm-thick periodically-poled KTP crystal inside a Sagnac interferometer. The photons generated in all the sources are filtered in wavelength and spatial mode by using narrow band interference filters and single-mode fibers, respectively. Then, the photons are shared pairwise among to three measurement stations, $A_1$, $A_2$ and $A_3$, performing separable measurements on the pairs of incoming photons.  Each measurement is performed in the polarization space of the photons,  through a quarter-waveplate (QWP), a half-waveplate (HWP), a polarizing beam splitter (PBS). In order to share photons along the stations in the triangle configuration, the photons are transmitted along single mode fibres (SMFs) with length up to $25$m.   
}
\label{fig:app}
\end{figure}

To experimentally implement the triangle network we exploit the photonic platform of Fig.~\ref{fig:app} with three different and separated polarization-entangled photon sources ($\rho_{12}, \rho_{23}, \rho_{13}$). The source $\rho_{12}$ generates photons by pumping a BBO crystal in pulsed mode, while the others, $\rho_{13}$ and $\rho_{23}$, are composed by ppKTP crystals pumped by a continuous wave laser. All three photon pairs exploit type-II degenerate spontaneous parametric down-conversion (SPDC) process.
All the generated photons are distributed among the laboratories, i.e., one remains in the laboratory where it was generated while the other one is sent to the adjoining laboratory through optical fibers, the maximum length of which was $25$ m. The three measurement stations $A_{1}, A_{2}$ and $A_{3}$ belonging to the laboratories 1, 2 and 3 respectively are composed by  half-wave plates (HWP), quarter-wave plates (QWP) and polarizing beam splitters (PBS). At both outputs of  each PBS, photons are coupled into single-mode fibers (SMF) and directed to single photons detectors (SPD). The electronic signals generated by the detectors are sent to three time-to-digital converters, each located in their respective measurement stations.

The independence of the generated states is ensured by the location of the sources and the distinct lasers used as pumps. In particular, the quantum source generating $\rho_{12}$ is supplied by a Ti:Sapphire mode-locked laser, with a repetition rate of $76$ MHz and wavelength of $397.5$ nm. This impulsed pump laser is focused on 2 mm-thick beta-barium borate (BBO) crystal. Instead, the quantum sources generating $\rho_{23}$ and $\rho_{13}$ are implemented exploiting two independent continuous-wave diode lasers characterized by a wavelength of $404$ nm and focused on distinct periodically-poled KTP crystals inside Sagnac interferometers. By pumping the crystal in the clockwise and anti-clockwise paths of the interferometer, both the signals can generate a photon pairs through a collinear type II SPDC process. Finally, recombining the two generations, in the dual wavelength PBS, the entangled photon pairs is obtained.

The three sources generate bipartite states, each of the form
\begin{equation}
\begin{split}
    \rho_{exp} &= v | \Psi^- \rangle \langle \Psi^- | + \\
    &+ (1-v) \left(  \frac{\lambda}{2} (| \Psi^+ \rangle \langle \Psi^+ | + | \Psi^- \rangle \langle \Psi^- |) + \frac{1-\lambda}{4} \mathbb{I} \right)
\end{split}
\label{eq:expstate}
\end{equation}
where $| \Psi^{\pm} \rangle = ( |10 \rangle \pm   |01 \rangle)/ \sqrt{2}$.
They are a mixture of the maximally entangled state with coloured and white noise.

\begin{figure}[t!]
\includegraphics[width=.99\columnwidth]{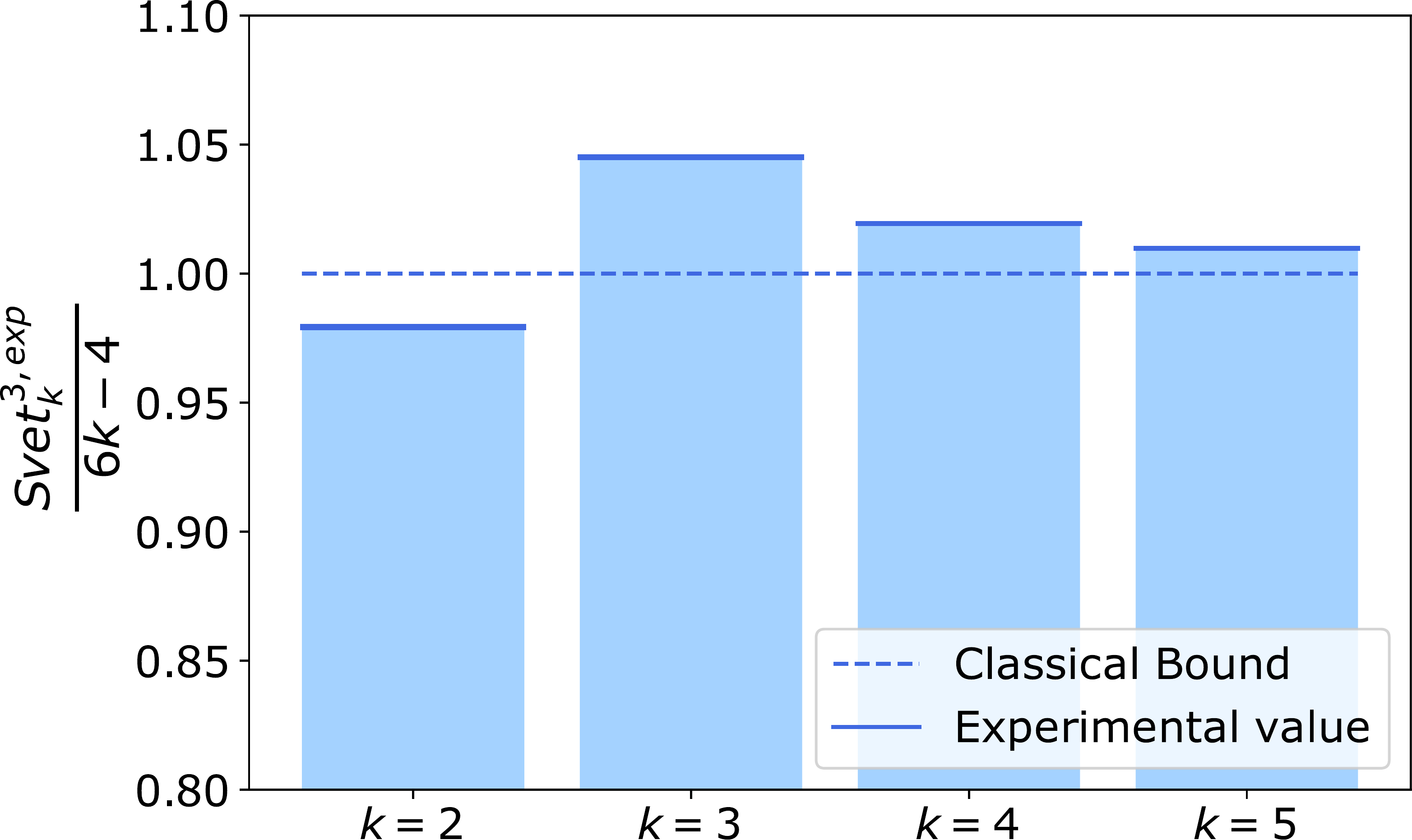}
\caption{\textbf{Experimental results showing genuine tripartite nonlocality in the triangle network with inputs.} 
The bars show the ratio $Svet^{3,exp}_k/(6k-4)$ with numbers of settings $k=2,3,4,5$ (notice that the inequality is not violated for $k=2$). Cyan bar charts represent the experimental data while the dashed lines represent the classical bounds for the nonlocal Svetlichny model \eqref{eq:svet_model1}. Error bars are calculated taking into account the Poissonian errors relative to the measured counts and are depicted by the blue upper parts of the bars.} 
\label{fig:3bell}
\end{figure}

To gauge the quality of the bipartite states we optimized the violation of the CHSH inequality \cite{clauser1969proposed}, corresponding to the chained inequality \eqref{eq:chained} with $k=2$ and being explicitly given by
\begin{equation}
        S_2^{ij} =  \left< A^i_1 A^j_1 \right> +\left< A^i_{2} A^j_{1} \right> + \left< A^i_2 A^j_2 \right> -\left< A^i_{1} A^j_{2} \right>  \leq 2.
    \label{eq:chsh}
\end{equation}
With our setup we have achieved the quantum violations of the classical bound given by $S_2^{12}= 2.643 \pm 0.002$, $S_2^{23}= 2.601 \pm 0.002$ and $S_2^{13}= 2.592 \pm 0.002$, values that unfortunately are not enough to violate the triangle SGMNL witness~\eqref{eq:svet_bi}. However, as we show next increase the number of measurements (increasing $k$) is enough to counteract the unavoidable effects of noise and experimental errors and violate the SGMNL witness~\eqref{eq:svet_bi}.

Experimentally, we have tested the triangle Svetlichny inequality \eqref{eq:svet_bi} on BKP games~\eqref{eq:chained}  with $k=2,3,4,5$ obtaining a violation only for $k\geq 3$. Hereafter we take $S^N_k\coloneqq \sum_{i=1}^N S_k^{i,i\oplus 1}$ where $\oplus$ denotes addition modulo $N$. More specifically, we obtain $Svet_2^{3,exp}=7.835 \pm 0.003$, $Svet_3^{3,exp}= 14.632 \pm 0.006$, $Svet_4^{3,exp}= 20.388 \pm 0.004$ and $Svet_5^{3,exp}= 26.252 \pm 0.006$. For $k\geq 3$  the bounds in \eqref{eq:svet_bi} are violated by $105$, $97$ and $45$ standard deviations respectively for $k=3,4$ and $5$. These results are summarized in Table \ref{tab}. The results are also presented in Fig.~\ref{fig:3bell} where we show the ratio $Svet_k^{3,exp}/(6k-4)$ (where $6k-4$ is the classical bound of \eqref{eq:svet_bi} applied to~\eqref{eq:chained}). As expected from the theoretical analysis the larger ratio is obtained for $k=3$. 

\begin{table}[htp]
\centering
\begin{tabular}{|c|c|c|c|}
\hline
k & Classical bound & Exp. value  & Ratio  \\
settings & $2B_L+B_s$ &  $Svet_k^{3,exp}$ & $Svet_k^{3,exp}/(6k-4)$ \\
\hline
\hline
$2$ & $8$ & $7.835 \pm 0.003$ & $0.9793 \pm 0.0004$ \\
$3$ &$14$ & $ 14.632 \pm 0.006$ & $1.0451 \pm 0.0004$\\
$4$ & $20$ & $20.388 \pm 0.004$ & $1.0194 \pm 0.0002$ \\
$5$ & $26$ & $26.252 \pm 0.006$ & $1.0098 \pm 0.0002$ \\
\hline
\end{tabular}
\caption{Considering the settings $k=2,3,4,5$, the second column shows the classical bound $2B_L+B_s=6k-4$, the third column what we achieve experimentally and the last column shows the ratio between the experimental value and the classical bound (a ratio larger than $1$ imply a quantum violation).}
\label{tab}
\end{table}

The data acquisition was performed using three separated time-to-digital converters (TDC), with a time resolution of $ \approx 81 $ ps, recording the timestamps of all measured events in each station. In particular, each node is provided by a TDC synchronized with each other using a shared random signal, acting as a time reference. The time of the three nodes is sent to a separated machine that supplies real-time synchronization via dedicated software. Subsequently, only the two-fold coincidence events detected between photons coming from the same source are considered. The coincidence window has been fixed equal to $ \approx 3.24 $ ns in order to limit the effect of the noisy measurements. If the three two-fold events, relative to each source, are all recorded in a given window, a six-fold event is extracted. Analyzing the violation for different values  of the time window, which has to be larger than that one used in the two-fold cases, the best value is selected by maximizing the violations. However, it is worth noticing that choosing a large window increases the number of 6-fold coincidences, improving the statistical error, but at the cost of introducing additional noise.

\begin{figure}[t!]
\includegraphics[width=1\columnwidth]{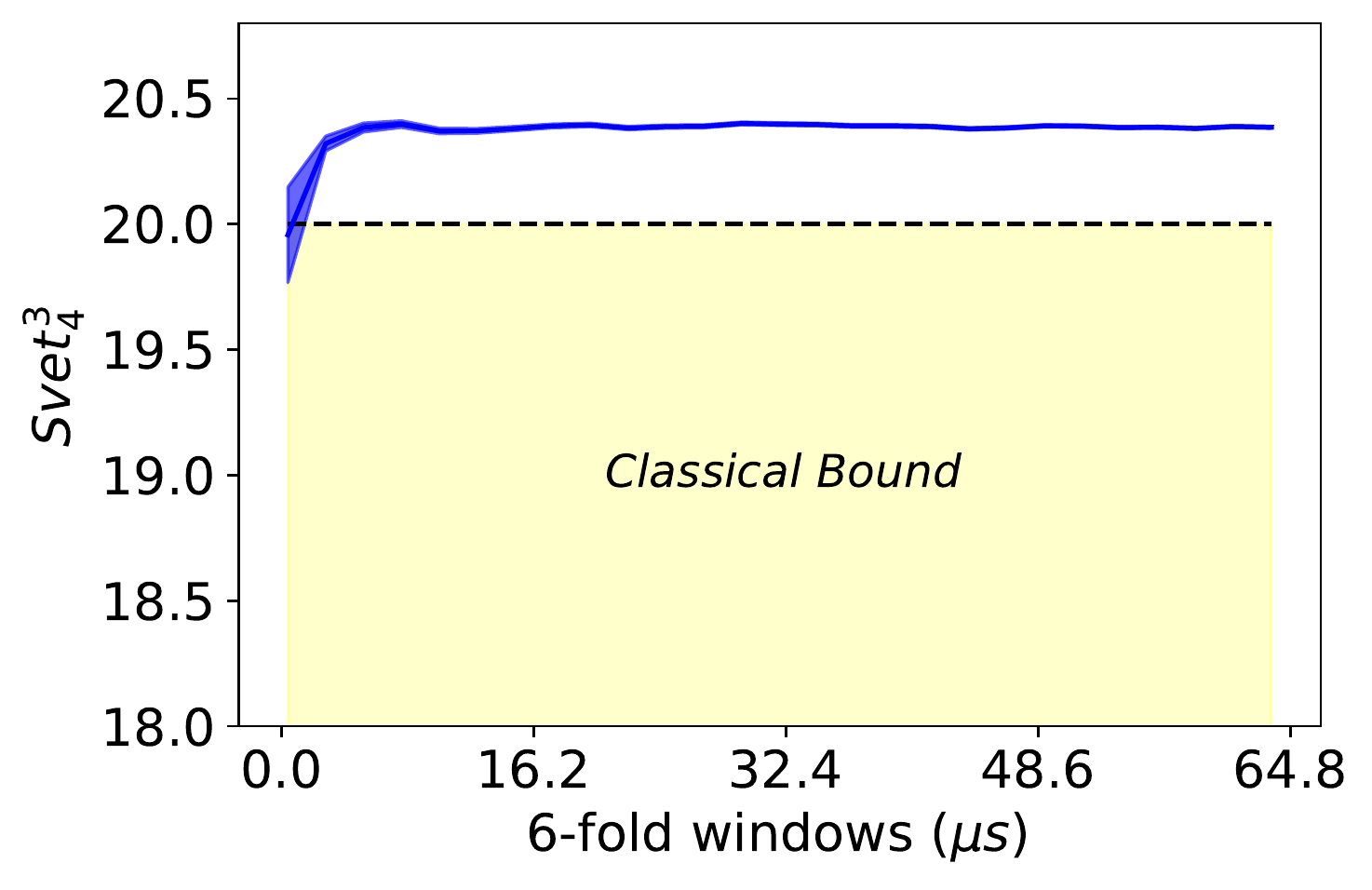}
\caption{\textbf{ Violation of $Svet_4^3$  as a function of the 6-fold coincidence window length} The blue line interpolates the experimental values of $Svet_4^3$ and the lighter blue region represents the relative Poissonian error.} 
\label{fig:timewindow}
\end{figure}

We performed all measurements with different $6$-fold coincidence windows, i.e. the time interval where detection events are considered simultaneous, ranging from $0.405$  to $63  \mu$s. Within this range, we chose a suitable trade-off between the width of the window and the 6-fold rate and consequently of the Poissonian uncertainty of the measurements. All results presented are obtained using a fixed window of $61 \mu $s. As an illustration of the role of these time windows, in Fig.~\ref{fig:timewindow} we show the obtained violation of the  SGMNL witness~\eqref{eq:svet_bi} as a function of this window.

\section{Discussion}
\label{sec:disc}
Historically, the primary interest in quantum networks has been the study of communication networks relevant to modelling the different scales of quantum internet \cite{Chen2021,wehner2018quantum,brito2020statistical}. More recently, however, quantum networks have emerged to take a central place in the foundations of quantum theory. This stems from an appreciation of that Bell's theorem --- the focus of the community for a long period of time ---  represents only a particular such network at the interface between quantum information and causality theory \cite{fritz2012beyond,henson2014theory,chaves2015information,steudel2015information,fritz2012beyond,fraser2018causal,wolfe2019inflation,krivachy2019neural,renou2019genuine,aaberg2020semidefinite,gisin2019entanglement,kraft2020quantum,vsupic2020quantum,baumer2021demonstrating,renou2019limits,branciard2012bilocal,chaves2014causal,chaves2016polynomial,tavakoli2021bilocal,tavakoli2014nonlocal,gisin2017all,poderini2019exclusivity,tavakoli2017correlations,rosset2016nonlinear,bancal2009,chaves2017causal,gisin2019constraints,canabarro2019machine}. Imposing different causal structures in quantum experiments allows us to probe the insufficiency of classical explanations in new regimes and thus to identify new --- and in some cases \emph{stronger} --- forms of nonclassical behaviour.

Within this context, our main goal was to demonstrate the possibility of achieving genuine multipartite nonlocality by having the parties merely perform local measurements on bipartite entanglement sources distributed in a network. Our theoretical analysis in Section~\ref{sec:sec3} here identified the tripartite triangle graph nonlocal game as the \emph{most} suitable configuration of parallel bipartite nonlocal games to robustly realize genuine multipartite nonlocality. The corresponding causal structure --- which formed the basis for our experimental implementation --- is the so-called triangle network, a causal scenario that has been at the focus of extensive theoretical analysis recently~\cite{henson2014theory,chaves2015information,steudel2015information,fritz2016beyond,fraser2018causal,wolfe2019inflation,krivachy2019neural,renou2019genuine,aaberg2020semidefinite,gisin2019entanglement,kraft2020quantum,vsupic2020quantum,baumer2021demonstrating,renou2019limits}. Whereas most prior literature has considered a form of triangle network without inputs to the parties' measurements and without shared randomness, we here consider more traditional multipartite nonlocal games. The salient common feature, however, is the independence of the three sources of quantum states. 

As expected, our experimental results are inconsistent with any local hidden variable model; that it, we achieved nonlocal correlations. Much more significantly, however, is that our experimental results are furthermore inconsistent with a \emph{stronger} classical causal hypothesis, namely, Svetlichny's classical foil theory which allows for any two of the three parties to communicate their inputs to coordinate their outputs with each other~\cite{svetlichny1987distinguishing}. 

Thus, employing our photonic setup we experimentally realized genuine tripartite nonlocality. Our operational witness of this fact was the violating of a class of Bell inequalities rederived here in Lemma~\ref{ref:dibound}. Such inequalities were initially considered in Ref.~\cite{chaves2017causal}; they consist of the parallel testing of the chained (a.k.a. BKP) Bell inequalities~\cite{braunstein1990wringing,BKP} between multiple pairs of distinct parties. Violation of such a combined multipartite inequality unambiguously witnesses Svetlichny-esque genuine multipartite nonlocality. 

To be clear, we are not claiming to be the first experiment to violate inequalities bounding the correlations consistent with Svetlichny's classical foil theory. Ref.~\cite{lavoie2009experimental}, for example, certainly did so already. What distinguishes our demonstration from prior ones, however, is that prior experimental realizations of Svetlichny-esque genuine tripartite nonlocality have relied on a single tripartite state of high fidelity with GHZ. By contrast, our demonstration required only \emph{bipartite} sources of entangled states. Notably, bipartite states can be easily prepared with higher coincidence rate
and fidelities with respect to the GHZ states  \cite{lavoie2009experimental,hamel2014direct, chaisson2021phase, Wang2016, Bouwmeester1999, Pan2000},
and thus offer a more scalable platform.

Importantly, Svetlichny's causal model has also found applications for the detection and quantification of multipartite entanglement \cite{bancal2009,moroder2013,barreiro2013demonstration} as well as in cryptographic applications such as secret sharing \cite{moreno2020device}. Notwithstading, all these applications had in mind a single source of correlations rather then the network we consider here. We believe that exploring the protocols and applications opened by the quantum triangle and its generalizations to growing networks is a promising venue for future research. An area for which we hope our results might provoke further developments.

\section*{Acknowledgements} 
We thank Robert Spekkens for the fruitful discussions. 
This work was supported by The John Templeton Foundation via The Quantum Information Structure of Spacetime (QISS) Project (qiss.fr) (the opinions expressed in this publication are those of the author(s) and do not necessarily reflect the views of the John Templeton Foundation)  Grant Agreement No.  61466, by MIUR via PRIN 2017 (Progetto di Ricerca di Interesse Nazionale): project QUSHIP (2017SRNBRK), by the Regione Lazio programme “Progetti di Gruppi di ricerca” legge Regionale n. 13/2008 (SINFONIA project, prot. n. 85-2017-15200) via LazioInnova spa and by the ERC Advanced Grant QU-BOSS
(Grant agreement no. 884676). This work was supported by the Serrapilheira Institute (grant
number Serra – 1708-15763). Research at Perimeter Institute is supported in part by the Government of Canada through the Department of Innovation, Science and Economic Development and by the Province of Ontario through the Ministry of Colleges and Universities. RC and AC also acknowledge the Brazilian National Council for Scientific and Technological Development (CNPq) via the National Institute for Science and Technology on Quantum Information (INCT-IQ) and Grants No. 406574/2018-9 and 307295/2020-6, the Brazilian agencies MCTIC and MEC. 


%\bibliographystyle{apsrev4-2-wolfe}
%\bibliography{mrefs}

\end{document}